\documentclass[aps,prl,twocolumn,groupedaddress]{revtex4-2}

\usepackage{amsmath,amsthm,amssymb,mathtools,tikz,booktabs,tensor,slashed,dutchcal,upgreek}
\usepackage[mathscr]{eucal}
\usetikzlibrary{cd}

\newcommand{\mc}{\mathcal} \newcommand{\mf}{\mathfrak} \newcommand{\mb}{\mathbb} \newcommand{\on}{\operatorname}  \newcommand{\ms}{\mathscr}
\newcommand{\slot}{\;\cdot\;} \newcommand{\la}{\langle} \newcommand{\ra}{\rangle}
\newcommand{\gm}{{\mc G}} \newcommand{\di}{\slashed D} \newcommand{\tr}{\on{Tr}}
\DeclareMathOperator{\gr}{\mathcal{R}}

\begin{document}


\title{Higher fermions in supergravity}



\author{Julian Kupka}
\email{j.kupka@herts.ac.uk}
\author{Charles Strickland-Constable}
\email{c.strickland-constable@herts.ac.uk}
\author{Fridrich Valach}
\email{f.valach@herts.ac.uk}
\affiliation{Department of Physics, Astronomy and Mathematics,
University of Hertfordshire, College Lane, Hatfield, AL10 9AB, United Kingdom}


\date{\today}

\begin{abstract}
We show that the generalised geometry formalism provides a new approach to the description of higher-fermion terms in $\ms N=1$ supergravity in ten dimensions, which does not appeal to supercovariantisation or superspace. We find expressions containing only five higher-fermion terms across the action and supersymmetry transformations, working in the second-order formalism.
\end{abstract}


\maketitle

\section{Introduction}
  Ten-dimensional $\ms N=1$ supergravity coupled to Yang--Mills multiplets \cite{Bergshoeff:1981um,Chapline:1982ww,Dine:1985rz} forms one of the basic cornerstones of string theory and has been the subject of many investigations over the span of several decades. Nevertheless, the vast majority of the treatments consider either the bosonic part of the action only or the action and supersymmetry transformations to the lowest non-trivial orders in fermions. Part of the reason is the immensely complicated structure of the four-fermion terms, obtained usually by reducing eleven-dimensional supergravity.
  
  In this letter we take a more direct approach using generalised geometry, which is naturally adapted to the left/right-moving worldsheet structure of string theory. Following the insights from \cite{pre}, we treat the spinor fields as half-densities, which leads to further significant simplifications. Using this language one can simply write down the most general admissible expression and then check its supersymmetry by hand. This leads to a surprisingly simple form of the action \eqref{a} and supersymmetry transformations \eqref{s}, extending the lower-order treatment in \cite{CSW1,Coimbra:2014qaa}. The generalised-geometric formulation is second-order throughout and makes manifest the compatibility of Poisson--Lie T-duality \cite{Klimcik:1995ux} with the supergravity equations of motion, extending the purely bosonic result of \cite{sv2}.
  
  This letter presents the results and the main details of the story, while the full calculation of supersymmetry of the action and the details of the reduction to classical variables is left for a later publication \cite{pre}.

\section{Generalised geometry}
  Generalised geometry, at least in the narrower original sense of the term, is the study of structures on Courant algebroids \cite{liu1997manin}. Up to small modifications (the most important being the treatment of fermions as half-densities) we will follow \cite{let,Coimbra:2014qaa}.
  
  We start with a Lie algebra $\mf g$ with an invariant pairing denoted by $\tr$, and take $M:=\mb R^{10}$. We then consider the bundle
  \begin{equation}
    E:=TM\oplus T^*M\oplus \on{ad} \ ,
  \end{equation}
  where $\on{ad}=M\times \mf g$ stands for the trivial bundle whose sections can be identified with $\mf g$-valued functions on $M$. The bundle $E$ is equipped with the following bracket on the space of its sections:
  \begin{equation}
  \begin{aligned}
        [x+\alpha+s,y+\beta+t]&=L_xy+(L_x\beta-i_yd\alpha+\tr t(ds))\\
        &\qquad+(i_xdt-i_yds+[s,t]_\mf g) \ ,
  \end{aligned}  
  \end{equation}
  as well as the pairing
  \begin{equation}
    \langle x+\alpha+s,y+\beta+t\rangle:=\alpha(y)+\beta(x)+\tr st \ ,
  \end{equation}
  and the \emph{anchor map} $a\colon E\to TM$,
  \begin{equation}
    a(x+\alpha+s):=x.
  \end{equation}
  
  One can also construct nontrivial ``global'' examples of Courant algebroids by gluing the above description on local patches together using automorphisms of the structure. 
  In this way one can describe field configurations involving non-trivial manifolds and topologically non-trivial field strengths. 
  Here we will suffice with the local picture.
  
  The bosonic field content of the supergravity will consist of two fields
  \begin{itemize}
    \item a half-density $\sigma$ on $M$
    \item a generalised metric $\gm$, seen as a symmetric endomorphism on $E$ such that $\gm^2=1$.
  \end{itemize} 
  Decomposing $E$ into eigenbundles of $\gm$ we get 
  \begin{equation} E=C_+\oplus C_- \ . \end{equation} 
  In order to make connection with supergravity, we shall from now on assume that
  \begin{itemize}
    \item the signature of $\la \slot,\slot\ra$ on $C_+$ is $(9,1)$
    \item $a(C_+)=TM$.
  \end{itemize}
  This allows us to recover the usual field content, since
  \begin{equation}
    C_+=\{x+(i_xg+i_xB-\tfrac12\tr A \,i_xA)+i_xA\mid x\in TM\}
  \end{equation}
  for some Lorentzian metric $g$, 2-form $B$, and a connection $A\in\Omega^1(M,\mf g)$. Similarly,
  \begin{equation}
    \sigma^2=\Phi=\sqrt{|g|}e^{-2\varphi},
  \end{equation}
  for a function $\varphi$ on $M$ (the dilaton). Note that the orthogonal complement $C_-=C_+^\perp$ can be written as the orthogonal direct sum of
  \begin{equation}
      \{x+(-i_xg+i_xB-\tfrac12\tr A\,i_xA)+i_xA\mid x\in TM\}
  \end{equation}
  and
  \begin{equation}
    \{0-\tr tA+t\mid t\in \on{ad}\}.
  \end{equation}
  
  To proceed, let us denote by $S_\pm$ the positive and negative chirality Majorana spinors w.r.t.\ $C_+$, and by $H$ the line bundle of half-densities on $M$ \footnote{Spinor densities have previously been considered in, e.g.,~\cite{Hillmann:2009ci,Gustafsson:1998ej,Choi:2022srv}.}. We shall use indices $A$, $a$, and $\alpha$ to label the frames of $E$, $C_+$, and $C_-$, respectively. The fermionic fields of the theory are
  \begin{itemize}
    \item a section $\rho$ of $S_+\otimes H$
    \item a section $\psi$ of $S_-\otimes C_-\otimes H$.
  \end{itemize}
  Since the anchor allows us to identify $C_+\cong TM$ and $C_-\cong TM\oplus \on{ad}$ and $\sigma$ allows the identification of half-densities with functions, we can recover the usual dilatino $\uprho$, gravitino $\uppsi$ and the gaugino $\upchi$ via
  \begin{equation}
    \rho \to \sqrt[4]2\sigma\uprho,\qquad \psi \to \sqrt[4]2\sigma\uppsi+\tfrac1{\sqrt[4]2}\sigma\upchi.
  \end{equation}
  
  Similarly to ordinary Riemannian geometry, one can define connections $D_uv$, where both $u$ and $v$ are sections of $E$, by demanding that for any function $f$ on $M$
  \begin{equation}
    D_{fu}v=fD_uv,\quad D_u(fv)=fD_uv+(a(u)f)v,
  \end{equation}
  and that $D$ preserves the pairing $\la\slot,\slot\ra$. Any such connection naturally acts also on half-densities via
  \begin{equation}
    D_u\sigma=L_{a(u)}\sigma-\tfrac12\sigma D_Au^A,
  \end{equation}
  where $L$ is the Lie derivative.  
  Given a pair $(\gm,\sigma)$ one can show \cite{Garcia-Fernandez:2016ofz} that there exists a connection $D$, which preserves both $\gm$ and $\sigma$ and has vanishing torsion tensor
  \begin{equation}
    T(u,v):=D_uv-D_vu-[u,v]+\la Du,v\ra.
  \end{equation}
  Such a connection is to be regarded as a natural analogue of the Levi-Civita connection; it is however not unique (there is an infinite class of Levi-Civita connections). Nevertheless, there are expressions that one can build out of $D$ which are only dependent on $\gm$ and $\sigma$ and not on the particular representative of the Levi-Civita class. The most important examples are
  \begin{equation}
    D_\alpha \rho,\quad D_\alpha\psi^\alpha, \quad \di \rho,\quad \di\psi^\alpha,
  \end{equation}
  with $\di:=\gamma^aD_a$, from which one can assemble the fermionic kinetic terms, and the generalised scalar curvature operator $\gr$
  which can be conveniently defined via
  \begin{equation}
    (\di^2+D^{\alpha}D_{\alpha})\lambda=-\tfrac18\mc R\lambda,
  \end{equation}
  for any spinor half-density $\lambda$ w.r.t.\ $C_+$.

  \section{Action and local supersymmetry}
    In terms of the above generalised-geometric ingredients the action for $\ms N=1$ supergravity in ten dimensions coupled to Yang--Mills multiplets takes the simple form
    \begin{equation}\label{a}
      \begin{aligned}
        S&=\smash{\int_M}\mc R\sigma^2+\bar\psi_{\alpha}\slashed D\psi^{\alpha}+\bar\rho\slashed D\rho+2\bar\rho D_{\alpha}\psi^{\alpha}\\
        &\qquad\qquad -\tfrac1{768}\sigma^{-2}(\bar\psi_{\alpha}\gamma_{abc}\psi^{\alpha})(\bar\rho\gamma^{abc}\rho)\\
        &\qquad\qquad-\tfrac1{384}\sigma^{-2}(\bar\psi_{\alpha}\gamma_{abc}\psi^{\alpha})(\bar\psi_{\beta}\gamma^{abc}\psi^{\beta}).
      \end{aligned}
    \end{equation}
    It can be shown by a direct calculation \cite{pre} (which can be fully exhibited on just a few pages) that this is invariant under the local supersymmetry transformations
    \begin{equation}\label{s}
    \begin{aligned}
        \delta \mc G_{ab}&=\delta\mc G_{\alpha\beta}=0,\quad \delta\mc G_{a\beta}=\delta\mc G_{\beta a}=\tfrac12\sigma^{-2}\bar \epsilon\gamma_a\psi_{\beta}\\
        \delta\sigma&=\tfrac18\sigma^{-1}(\bar\rho\epsilon)\\
        \delta\rho&=\slashed D\epsilon+\tfrac1{192}\sigma^{-2}(\bar\psi_{\alpha}\gamma_{abc}\psi^{\alpha})\gamma^{abc}\epsilon\\
        \delta\psi_{\alpha}&=D_{\alpha}\epsilon+\tfrac18\sigma^{-2}(\bar\psi_{\alpha}\rho)\epsilon+\tfrac18\sigma^{-2}(\bar\psi_{\alpha}\gamma_a\epsilon)\gamma^a\rho
    \end{aligned}
    \end{equation}
    where $\epsilon$ is an (odd) section of $S_-\otimes H$ \footnote{See also the work~\cite{Baron:2024tph} which appeared shortly after ours, and which conjectured a similar general form for the quartic fermion terms.}.
    
    Using the aforementioned decomposition
    \begin{equation}
      \gm \to (g,B,A),\quad\sigma\to \varphi,\quad \rho\to\uprho,\quad \psi\to(\uppsi,\upchi)
    \end{equation}
    the above formulae become
    \begin{equation}
    \label{long-A}
    \begin{aligned}
    S=\!\smash{\int_M}\!&\Phi(R+4|\nabla \varphi|^2-\tfrac1{12}H_{\mu\nu\rho}H^{\mu\nu\rho}\\
    &\;\quad+\tfrac14\tr F_{\mu\nu}F^{\mu\nu}+\tfrac12\tr\bar\upchi\slashed\nabla_{\hspace{-1mm}A}\upchi\\
    &\;\quad-\bar\uppsi^\mu\slashed\nabla\uppsi_\mu+\uprho\slashed\nabla\uprho-2\bar\uppsi^\mu\nabla_\mu\uprho\\
    &\;\quad+\tfrac14\bar\uppsi^\mu\slashed H\uppsi_\mu-\tfrac14\bar\uprho\slashed H\uprho-\tfrac18\tr\bar\upchi\slashed H\upchi\\
    &\;\quad+\tfrac12 H_{\mu\nu\rho}\bar\uppsi^\mu \gamma^\nu\uppsi^\rho+\tfrac14\bar\uppsi^\mu H_{\mu\nu\rho}\gamma^{\nu\rho}\uprho\\
    &\;\quad+\tfrac12\tr\bar\upchi\slashed F\uprho+\tr F_{\mu\nu}\bar\uppsi^\mu \gamma^\nu\upchi\\
    &\;\quad+\tfrac1{384}(\bar\uppsi_{\mu}\gamma_{\nu\rho\sigma}\uppsi^{\mu})(\bar\uprho\gamma^{\nu\rho\sigma}\uprho)\\
    &\;\quad-\tfrac1{768}(\bar\uprho\gamma^{\mu\nu\rho}\uprho)\tr(\bar\upchi\gamma_{\mu\nu\rho}\upchi)\\
    &\;\quad-\tfrac1{192}(\bar\uppsi_{\mu}\gamma_{\rho\sigma\tau}\uppsi^{\mu})(\bar\uppsi_{\nu}\gamma^{\rho\sigma\tau}\uppsi^{\nu})\\
    &\;\quad+\tfrac1{192}(\bar\uppsi_{\mu}\gamma_{\nu\rho\sigma}\uppsi^{\mu})\tr(\bar\upchi\gamma^{\nu\rho\sigma}\upchi)\\
    &\;\quad-\tfrac1{768}\tr(\bar\upchi\gamma_{\mu\nu\rho}\upchi)\tr(\bar\upchi\gamma^{\mu\nu\rho}\upchi)),
  \end{aligned}
  \end{equation}
  and
  \begin{equation}
  \label{long-S}
    \begin{aligned}
      \delta g_{\mu\nu}&=\bar\upepsilon\gamma_{(\mu}\uppsi_{\nu)}\\
      \delta B_{\mu\nu}&=\bar\upepsilon\gamma_{[\mu}\uppsi_{\nu]}-\tr A_{[\mu}\bar\upepsilon\gamma_{\nu]}\upchi\\
      \delta A_\mu&=-\tfrac12\bar\upepsilon\gamma_\mu\upchi\\
      \delta \varphi&=\tfrac14\bar\uprho\upepsilon-\tfrac14\bar\uppsi^\mu\gamma_\mu\upepsilon\\
      \delta \uprho&=-\slashed\nabla \upepsilon+(\nabla_\mu\varphi)\gamma^\mu\upepsilon+\tfrac14\slashed H\upepsilon\\
      &\;\quad+\tfrac1{96}(\bar\uppsi_\mu \gamma_{\nu\rho\sigma}\uppsi^\mu)\gamma^{\nu\rho\sigma}\upepsilon+\tfrac14(\bar\uprho\upepsilon)\uprho\\
      &\;\quad-\tfrac1{192}\tr(\bar\upchi\gamma_{\mu\nu\rho}\upchi)\gamma^{\mu\nu\rho}\upepsilon\\
      \delta\uppsi_\mu&=\nabla_\mu\upepsilon-\tfrac18H_{\mu\nu\rho}\gamma^{\nu\rho}\upepsilon\\
      &\;\quad-\tfrac14(\bar\uppsi_\mu\uprho)\upepsilon-\tfrac14(\bar\uppsi_\mu\gamma_\nu\upepsilon)\gamma^\nu\uprho+\tfrac14(\bar\uprho\upepsilon)\uppsi_\mu\\
      \delta\upchi&=\tfrac12\slashed F\upepsilon-\tfrac14(\bar\upchi\uprho)\upepsilon-\tfrac14(\bar\upchi\gamma_\mu\upepsilon)\gamma^\mu\uprho+\tfrac14(\bar\uprho\upepsilon)\upchi,
    \end{aligned}
    \end{equation}
    where
    \begin{equation}
    \begin{aligned}
      H&:=dB+\tfrac12\tr(A\wedge dA)+\tfrac16\tr(A\wedge[A,A])\\
      \slashed\nabla_{\!\!A}\upchi&:=\slashed \nabla \upchi+\gamma^\mu[A_\mu,\upchi]_\mf g,
    \end{aligned}
    \end{equation}
    and $\slashed C:=\tfrac1{p!}\gamma^{\mu\dots\nu}C_{\mu\dots\nu}$ for a $p$-form $C$.    
    Our definition of the dilatino relates to the other common definition (typically denoted $\uplambda$) via
    \begin{equation}
      \uplambda=\gamma_\mu\uppsi^\mu+\rho.
    \end{equation}
    Although significantly longer, the expressions~\eqref{long-A} and~\eqref{long-S} still provide simplifications when compared to the standard ones.
    The fact that this has to coincide --- up to simple field redefinitions and Fierz identities --- with the other known expressions for 10-dimensional supergravity (e.g.\ the ones obtained by reducing the 11-dimensional theory) follows from the fact that the expressions match up to the lowest nontrivial order in fermions \cite{Bergshoeff:1981um,Chapline:1982ww,Dine:1985rz,CSW1,Coimbra:2014qaa} and from the uniqueness of the supergravity theory.
  
  \section{Discussion}
    There are several conclusions that can be drawn from this generalised-geometric reformulation.
    
    First, the simplicity of the expressions confirms that generalised geometry provides a natural set of variables for studying the massless sector of string theory. 
    Further, it has enabled us to find compact expressions for the higher-fermion terms which do not involve the usual constructions of supercovariant quantities or superspace. 
    
    Second, it is easy to see that \eqref{a} and \eqref{s} remain meaningful even in the case of a general Courant algebroid, as long as one demands that the signature of $C_+$ is $(9,1)$, $(5,5)$, or $(1,9)$ \footnote{Strictly speaking one should also require that $\on{rank}(C_-)\neq 1$ and that the bundle $C_+$ is spin, i.e.\ it admits spinors.}; in particular one can take $\dim M$ to be arbitrary.
     This means that one can study various interesting ``limits'' of the theory. For instance, letting $C_+=E$ one obtains a locally supersymmetric topological theory whose only fields are $\sigma$ and $\rho$ \cite{Kupka:2024tic}. Similarly, we can take $M$ to be a point, in which case the Courant algebroid becomes simply a Lie algebra with an invariant pairing. The expressions \eqref{a} and \eqref{s} then describe a theory with finitely many degrees of freedom, which nevertheless still retains some of the aspects of the 10-dimensional supergravity, in particular its symmetry structure, and thus provides an interesting toy model.

    Finally, since Poisson--Lie T-duality \cite{Klimcik:1995ux} can be formulated in terms of pullbacks of Courant algebroids \cite{Severa:2016lwc,sv2}, a repetition of the arguments in \cite{sv2} shows that this duality is compatible with the equations of motion of the full supergravity.

\appendix

\begin{acknowledgments}
C.S.-C.~and F.V.~are supported by an EPSRC New Investigator Award, grant number EP/X014959/1. No new data was collected or generated during the course of this research. 
\end{acknowledgments}

\bibliography{citations}

\end{document}